\documentclass[aps,prl,reprint]{revtex4-1}

\usepackage{amsmath}
\usepackage{dcolumn}
\usepackage{graphicx}
\usepackage{verbatim}
\usepackage[colorlinks, linkcolor=blue, citecolor=blue, urlcolor=blue]{hyperref}

\newcommand{\f}{\frac}
\newcommand{\ep}{\epsilon}
\newcommand{\om}{\omega}
\newcommand{\Mpl}{M_{\scriptscriptstyle\rm  Pl}}
\newcommand{\ph}{\phi_{\scriptscriptstyle\rm VEV}}
\newcommand{\W}{\scriptscriptstyle\rm WD}
\newcommand{\N}{\scriptscriptstyle\rm NS}
\newcommand{\sun}{\scriptscriptstyle\odot}

\begin{document}
\title{Evidence of deviations from general relativity in binary pulsars?}

\author{Xing Zhang$^{1,2}$}\email{starzhx@mail.ustc.edu.cn}
\author{Wen Zhao$^{1,2}$}\email{wzhao7@ustc.edu.cn}
\author{Tan Liu$^{1,2}$}
\author{Kai Lin$^{3,4}$}
\author{Chao Zhang$^5$}
\author{Shaojun Zhang$^3$}
\author{Xiang Zhao$^5$}
\author{Tao Zhu$^3$}
\author{Anzhong Wang$^{3,5}$}\email{Anzhong$_$Wang@baylor.edu}

\affiliation{$^1$CAS Key Laboratory for Researches in Galaxies and Cosmology, Department of Astronomy, \\ University of Science and Technology of China, Chinese Academy of Sciences, Hefei, Anhui 230026, China}
\affiliation{$^2$School of Astronomy and Space Science, University of Science and Technology of China, Hefei 230026, China}
\affiliation{$^3$Institute for Advanced Physics $\&$ Mathematics, Zhejiang University of Technology, Hangzhou 310032, China}
\affiliation{$^4$Universidade Federal de Itajub\'a, Instituto de F\'isica e Qu\'imica, Itajub\'a, MG, Brasil}
\affiliation{$^5$GCAP-CASPER, Physics Department, Baylor University, Waco, TX 76798-7316, USA}

%%%%%%%%
\begin{abstract}
Testing gravitational theories by binary pulsars nowadays becomes a key issue. For the general screened modified gravity (SMG), the post-Keplerian parameters in the neutron star (NS) - white dwarf (WD) binaries differ from those of general relativity (GR), and the differences are quantified by the scalar charge $\epsilon_{\W}$ of WD. After deriving the constraints on $\epsilon_{\W}$ from four different NS-WD binaries, we find that $\epsilon_{\W}$ is different from zero at the 2$\sigma$ level in all the cases studied, and there exists an inverse correlation between masses and scalar charges of low-mass WDs, which is consistent with the screening mechanisms. In particular, two independent binaries with measured radii of WDs follow the coincident constraints on the vacuum expectation value of the scalar field. These self-consistent results indicate that the observations in NS-WD binary pulsars seem in favor of SMG, rather than GR.
\end{abstract}

\maketitle

%%%%%%%%%%%%%%%%%%%%%%%%%1
{\emph{Introduction.}}---Although Einstein's general relativity (GR) is one of the most successful theories of gravity, it suffers from the quantization, singularity, as well as dark matter and dark energy problems. For these reasons, testing GR in various circumstance is still the key task in modern physics \cite{Will1993p}. However, scientists can never truly prove that a theory (including GR) is correct, but rather all we do is to disprove, or more accurately to constrain alternative hypothesis \cite{Popper2002p}. Therefore, studies of modified theories of gravity play a crucial role in testing GR. As the minimal extension of GR, scalar-tensor theories are a natural alternative \cite{Fujii2003p,Damour1992p2093}, which invoke a conformal coupling between matter and an underlying scalar field. Meanwhile, the screening mechanisms are needed in order to evade the tight constraints of the theories from the Solar System and laboratories, which include the chameleon \cite{Khoury2004p44026,Khoury2004p171104}, symmetron \cite{Hinterbichler2010p231301,Hinterbichler2011p103521} and dilaton \cite{Damour1994p532,Brax2010p63519} mechanisms. These theories can be described within a unified theoretical framework called the screened modified gravity (SMG) \cite{Brax2012p44015}.

Thanks to the accurate measurements of various post-Keplerian (PK) parameters, since the discovery of the Hulse-Taylor binary, binary pulsars become the excellent laboratories for testing gravitational theories in the strong field regime \cite{stairs2003,ST-pulsar,other-pulsar}. For SMG, the neutron star (NS) - white dwarf (WD) binaries are the ideal targets to probe deviations from GR, since the extra scalar dipole radiations can be emitted by the systems. In the previous work \cite{Zhang2017p104027}, we constrained SMG by the observed rate of orbital period decay of the binary system PSR J1738+0333. As a comprehensive extension, in this Letter we consider the full PK parameters of the NS-WD binary, and find that the deviations from GR can be elegantly quantified by a unique parameter, the scalar charge of the WD involved. Analyzing four such independent systems, we find that the scalar charge deviates from zero in all of these cases, and the data show a tendency of the inverse correlation between masses and scalar charges of WDs, which is consistent with the prediction of the screening mechanisms. In particular, we find that two independent constraints from PSRs J1738$+$0333 and J0348$+$0432 with measured radii of WDs coincide with each other. These evidences are in favor of SMG, and represent slight (but yet consistent) deviations from GR.

%%%%%%%%%%%%%%%%%%%%%%%%%2
{\emph{Screened modified gravity.}}---The Lagrangian density of the most general SMG can be written as \cite{Damour1992p2093,Brax2012p44015},
\[
\mathcal{L}=\sqrt{-g}\Big[\frac{\Mpl^2}{2}R-\f{(\partial\phi)^2}{2}-V(\phi)\Big]+\mathcal{L}_m\big(A^2(\phi) g_{\mu\nu},\psi_m\big),
\]
where $\Mpl$ is the reduced Planck mass, and $V(\phi)$ is the bare potential of the scalar field $\phi$. The conformal coupling function $A(\phi)$ characterizes the interaction between $\phi$ and matter fields, collectively denoted by $\psi_m$, which induces the fifth force.

In order that SMG can generate a screening effect to suppress this fifth force in high density environments, the effective potential of the scalar field must have a minimum, acting as the physical vacuum \cite{Brax2012p44015}. Around this vacuum, the scalar field acquires an effective mass, which increases as the ambient density increases. Therefore, the scalar field can be screened and evades constraints in high density regions, while in low density regions, the long-range fifth force may affect galactic dynamics \cite{Gronke2015p123,Schmidt2010p103002}. In addition, the scalar field can also act as dark energy to provide the late-time acceleration of the Universe \cite{Khoury2004p44026,Hinterbichler2011p103521}. Meanwhile, the tensor gravitational waves (GWs) in SMG contain two basic polarization modes and all propagate with the speed of light \cite{liu2018}, whereby the severe constraints on the speeds of GWs obtained from GW170817 are satisfied \cite{Abbott2017p13,Abbott2017p161101}.

%%%%%%%%%%%%%%%%%%%%%%%%%3
{\emph{Post-Keplerian parameters.}}---In binary pulsars, PK parameters describe the relativistic corrections to the Keplerian orbit and provide excellent windows to test theories of gravity \cite{Damour1986p263, Damour1992p1840}. For a binary pulsar in a quasi-elliptical orbit with pulsar and companion masses $m_p$ and $m_c$, the five PK parameters in SMG are derived in \cite{Zhang2017p104027,Zhangp}, which are the average rate $\dot{\om}$ of the periastron advance, the amplitude $\gamma$ of the Einstein delay, the range $r$ and shape $s$ of the Shapiro delay, and the average rate $\dot{P_b}$ of the orbital period decay are given, respectively, by,
\begin{eqnarray}
\dot{\om}&=&3\left(\f{P_{b}}{2\pi}\right)^{-\f{5}{3}}\f{(T_{\sun}m)^{\f{2}{3}}}{1-e^2}\f{24+8\ep_p\ep_c-\ep^2_p\ep^2_c}{24(1+\ep_p\ep_c/2)^{\f{4}{3}}}, \\
\gamma&=&e\left(\f{P_b}{2\pi}\right)^{\f{1}{3}}T_{\sun}^{\f{2}{3}}\f{m_c}{m^{\f{1}{3}}}\Big(1+\f{m_c}{m}\Big)\Big(1+\f{1}{2}\ep_p\ep_c\Big)^{\f{2}{3}},
\\
r&=&T_{\sun}{m_c}, \\
s&=&x_p\left(\f{{P_b}}{2\pi}\right)^{-\f{2}{3}}T_{\sun}^{-\f{1}{3}}\f{m^{\f{2}{3}}}{m_c}\Big(1+\f{1}{2}\ep_p\ep_c\Big)^{-\f{1}{3}}, \\
\dot{P_b}&=&-2\pi\left(\f{P_b}{2\pi}\right)^{\!-\f{5}{3}}\!\f{T_{\sun}^{\f{5}{3}}m_pm_c}{m^{\f13}(1-e^2)^{\f72}}
\bigg\{\f{96}{5}\Big(1\!+\!\f{73e^2}{24}\!+\!\f{37e^4}{96}\Big) \nonumber
\\&&
+\!\left(\!\f{P_b}{2{\pi}T_{\sun}m}\!\right)^{\!\f23}\!\f{\ep_d^2}{2}\Big(1\!-\!\f{e^2}{2}\!-\!\f{e^4}{2}\Big)\!+\!\Big(8\ep_p\ep_c\!-\!\f{m_pm_c}{m^2}\ep_d^2\Big)
\nonumber\\&&
+\f{e^2}{12}\Big[335\ep_p\ep_c+\Big(9-24\f{m_pm_c}{m^2}\Big)\ep_d^2+21\Gamma^2\Big]
\nonumber\\&&
+\f{e^4}{48}\Big[191\ep_p\ep_c+\Big(9-6\f{m_pm_c}{m^2}\Big)\ep_d^2+21\Gamma^2\Big]
\bigg\},\label{P_decay2}
\end{eqnarray}
where masses are expressed in solar units, $T_{\sun}\equiv G M_{\sun}= 4.925490947 \mu \rm s$ is a solar mass in time units, $m\equiv m_p+m_c$ is the total mass, and $P_b$, $e$ and $x_p$ are the orbital period, orbital eccentricity and projected semimajor axis of the pulsar orbit, respectively. The quantities $\ep_p$ and $\ep_c$ are the pulsar and companion scalar charges. The other quantities are defined as $\ep_d\equiv \ep_c-\ep_p$ and $\Gamma\equiv \ep_cm_p/m+\ep_pm_c/m$. Note that the PK parameters depend on the masses and scalar charges of the pulsar and its companion, and reduce to the results of GR when $\ep_p=\ep_c=0$.

In Eq. \eqref{P_decay2}, the first and second terms are the tensor quadrupole and scalar dipole radiations, respectively, and the remaining terms represent the contributions from the monopole and the monopole-quadrupole and dipole-octupole cross terms. Clearly, the dipole radiation dominates the orbital period decay unless $\ep_c-\ep_p=0$, because $P_b/T_{\sun}=\mathcal{O}(10^{9})$ for a typical binary pulsar with a 1-hour orbital period.

%%%%%%%%%%%%%%%%%%%%%%%%%4
{\emph{Scalar charge.}}---The scalar charge (i.e., the screened parameter) in SMG is equivalent to the sensitivity \cite{Eardley1975p59a,Zhang2017p104027}, which characterizes how the gravitational binding energy of the object responds to its motion relative to the scalar field. The screening mechanisms imply that the scalar charge should be smaller for more compact objects. This is completely different from other alternative theories of gravity without screening mechanisms, which generally predict the large non-GR effects for compact objects \cite{Will1993p}. We consider the star (labeled as $a$) approximately as a uniform density sphere, and then the scalar charge is given by \cite{Zhang2016p124003}
\begin{align}
\label{epsilon_a}
\ep_a=\frac{\ph-\phi_a}{\Mpl\Phi_{a}},
\end{align}
where $\Phi_a \equiv Gm_a/R_a$ is the compactness of the star, $\phi_a$ is the position of the effective potential minimum inside the star, and $\ph$ is the vacuum expectation value (VEV) of the scalar field which depends on the background matter density. Note that in general  $\phi_a$ is inversely correlated to the matter density $\rho$ \cite{Zhang2016p124003}. Since the background matter density is always much less than that of compact stars, we have $\ph\gg\phi_{a}$. Obviously, the scalar charge is inversely proportional to the compactness, which agrees with the screening mechanisms.

%%%%%%%%%%%%%%%%%%%%%%%%%5
{\emph{Binary pulsars.}}---The scalar dipole radiation dominates the orbital period decay and depends on the difference in the scalar charges, hence the asymmetric systems like the NS-WD binary pulsars are one of the ideal targets for testing theories of gravity. Due to the screening mechanisms (i.e.~$\ep_{\W}/\ep_{\N}\sim10^{4}$), we can set $\ep_{\N}=0$ (compared with $\ep_{\W}$). Thus, the PK parameters ($\dot{\omega}$, $\gamma$, $s$) reduce to those of GR, and the constraint of $\ep_{\W}$ comes only from $\dot{{P}_b}$ \cite{footnote1}. To determinate the system parameters ($m_{\N}$, $m_{\W}$,  $\ep_{\W}$), we need at least three observables, including the intrinsic value of $\dot{{P}_b}$. With these requirements, we consider the four NS-WD systems: PSRs J1141$-$6545 \cite{Bhat2008p124017,Ord2002p75}, J1738$+$0333 \cite{Freire2012p3328a}, J0348$+$0432 \cite{Antoniadis2013p6131} and J1012$+$5307 \cite{Lazaridis2009p805,Desvignes2016p3341a}. The relevant parameters for these systems are listed in Table \ref{tab1_PSRs}.

%-----------------Tab1-----------------%
\begin{table*}[htbp]
\centering
\caption{Timing model parameters for four binary pulsar systems. Numbers in parentheses represent 1$\sigma$ (68.3\%) uncertainties in the last quoted digit. $^{\rm a}$The masses are derived by assuming that GR is valid.}
\label{tab1_PSRs}
\renewcommand{\arraystretch}{1.1}
\begin{tabular}{lrrrr}
\hline\hline
PSR Name & J1141$-$6545 \cite{Bhat2008p124017,Ord2002p75}
& J1738$+$0333 \cite{Freire2012p3328a}
& J0348$+$0432 \cite{Antoniadis2013p6131}
&J1012$+$5307 \cite{Lazaridis2009p805,Desvignes2016p3341a}
\\ \hline
Orbital period, $P_b$ (days)&0.1976509593(1)&0.3547907398724(13)&0.102424062722(7)&0.60467271355(3)\\
Projected semimajor axis, $x_p$ (s)&1.858922(6)&0.343429130(17)&0.14097938(7)&0.5818172(2)\\
Eccentricity, $e$&0.171884(2)&$0.34(11)\times10^{-6}$&$0.24(10)\times10^{-5}$&$1.2(3)\times10^{-6}$\\
Periastron advance, $\dot\omega$ (deg/yr)&5.3096(4)& ... & ... & ... \\
Einstein delay, $\gamma$ (ms)&0.773(11)& ... & ... &...  \\
Observed $\dot P_b$, $\dot P_b^{\text{obs}}$ $(10^{-13})$&$-4.03(25)$&$-0.170(31)$&$-2.73(45)$&$0.61(4)$\\
Intrinsic $\dot P_b$, $\dot P_b^{\text{int}}$ $(10^{-13})$&$-4.01(25)$&$-0.259(32)$&$-2.73(45)$&$-0.29(21)$\\
Shapiro delay, $s$ &0.97(1)& ... & ... & ... \\
Mass ratio, $q=m_{\N}/m_{\W}$& ... &8.1(2)&11.70(13)&10.5(5) \\
Pulsar mass, $m_{\N}$ ($M_{\sun}$)&$1.27(1)^{\rm a}$&${1.46^{+0.06}_{-0.05}}^{\rm a}$&$2.01(4)^{\rm a}$&$1.64(22)^{\rm a}$ \\
WD mass, $m_{\W}$ ($M_{\sun}$)&$1.02(1)^{\rm a}$&${0.181^{+0.008}_{-0.007}}$&$0.172(3)$&$0.16(2)$\\
WD radius, $R_{\W}$ ($R_{\sun}$)& ... &$0.037_{-0.003}^{+0.004}$&$0.065(5)$& ... \\
\hline\hline
\end{tabular}
\end{table*}
%-----------------Tab1-----------------%

%%%%%%%%%%%%%%%%%%%%%%%%%6
{\emph{Method and results.}}---We set up Monte Carlo simulations to determine the system parameters ($m_{\N}$, $m_{\W}$,  $\ep_{\W}$) for the four systems mentioned above, and the results are summarized in Table \ref{tab2_results}. In the simulations, all observables are randomly sampled from a normal distribution with mean and standard deviation equal to their fitted values and uncertainties, respectively, and then the unknown quantities can be obtained by solving the PK equations. This process is repeated $10^5$ times to construct the histograms of the system parameters and determine their median values and uncertainties.

%-----------------Tab2-----------------%
\begin{table}[htbp]
\centering
\caption{Parameters ($m_{\N}$, $m_{\W}$, $\ep_{\W}$) for binary pulsars.}
\label{tab2_results}
\renewcommand{\arraystretch}{1.1}
\begin{tabular}{p{1.7cm}<{\raggedright}p{1.1cm}<{\centering}p{1.9cm}<{\raggedright}p{1.5cm}<{\raggedright}p{1.4cm}<{\raggedleft}}
\hline\hline
PSR Name &CL(\%)& $\,\,m_{\N}(M_{\sun})$ & $m_{\W}(M_{\sun})$ & {$\ep_{\W}(\!10^{\!-3}\!)$} \\
\hline
J1141$-$6545 &$68.3$  &$1.273_{-0.011}^{+0.011}$ &$1.016_{-0.011}^{+0.011}$ & {$2.6_{-1.1}^{+1.0}$} \\
&$95.4$  & $1.273_{-0.022}^{+0.022}$ & $1.016_{-0.022}^{+0.022}$ &  {$2.6_{-2.0}^{+1.8}$} \\[0.6mm]
J1738$+$0333 &$68.3$   &$1.431^{+0.063}_{-0.063}$ & $0.177^{+0.007}_{-0.007}$ &  { $2.0_{-0.9}^{+1.0}$}  \\
&$95.4$   &$1.431^{+0.126}_{-0.125}$ &  $0.177^{+0.013}_{-0.014}$ &  { $2.0_{-1.6}^{+1.9}$}  \\[0.6mm]
{J0348$+$0432} &$68.3$  &$2.008_{-0.041}^{+0.041}$ & $0.172_{-0.003}^{+0.003}$&  {$4.6^{+1.9}_{-2.0}$} \\
&$95.4$  & $2.008_{-0.082}^{+0.083}$ & $0.172_{-0.006}^{+0.006}$  &  {$4.6^{+3.5}_{-3.6}$} \\[0.6mm]
{J1012$+$5307} &$68.3$  & $1.67^{+0.23}_{-0.22}$ & $0.16^{+0.02}_{-0.02}$ &  {$9.3^{+3.4}_{-3.7}$} \\
&$95.4$  & $1.67^{+0.46}_{-0.43}$ & $0.16^{+0.04}_{-0.04}$ &  {$9.3^{+6.4}_{-7.0}$} \\
\hline\hline
\end{tabular}
\end{table}
%-----------------Tab2-----------------%

We present our analysis of PSR J1141$-$6545 as an example. This system provides four PK observables ($\dot\omega$, $\gamma$, $\dot{P}_b$, $s$). Utilizing the first three, the system masses and WD scalar charge are derived and listed in Table \ref{tab2_results}. These masses imply the Shapiro delay shape $s=0.959\pm0.008$ at 68.3\%~confidence level (CL), which agrees with its observed value (see Table \ref{tab1_PSRs}).

\begin{figure}[htbp]
\centering
\includegraphics[width=7cm, height=7cm]{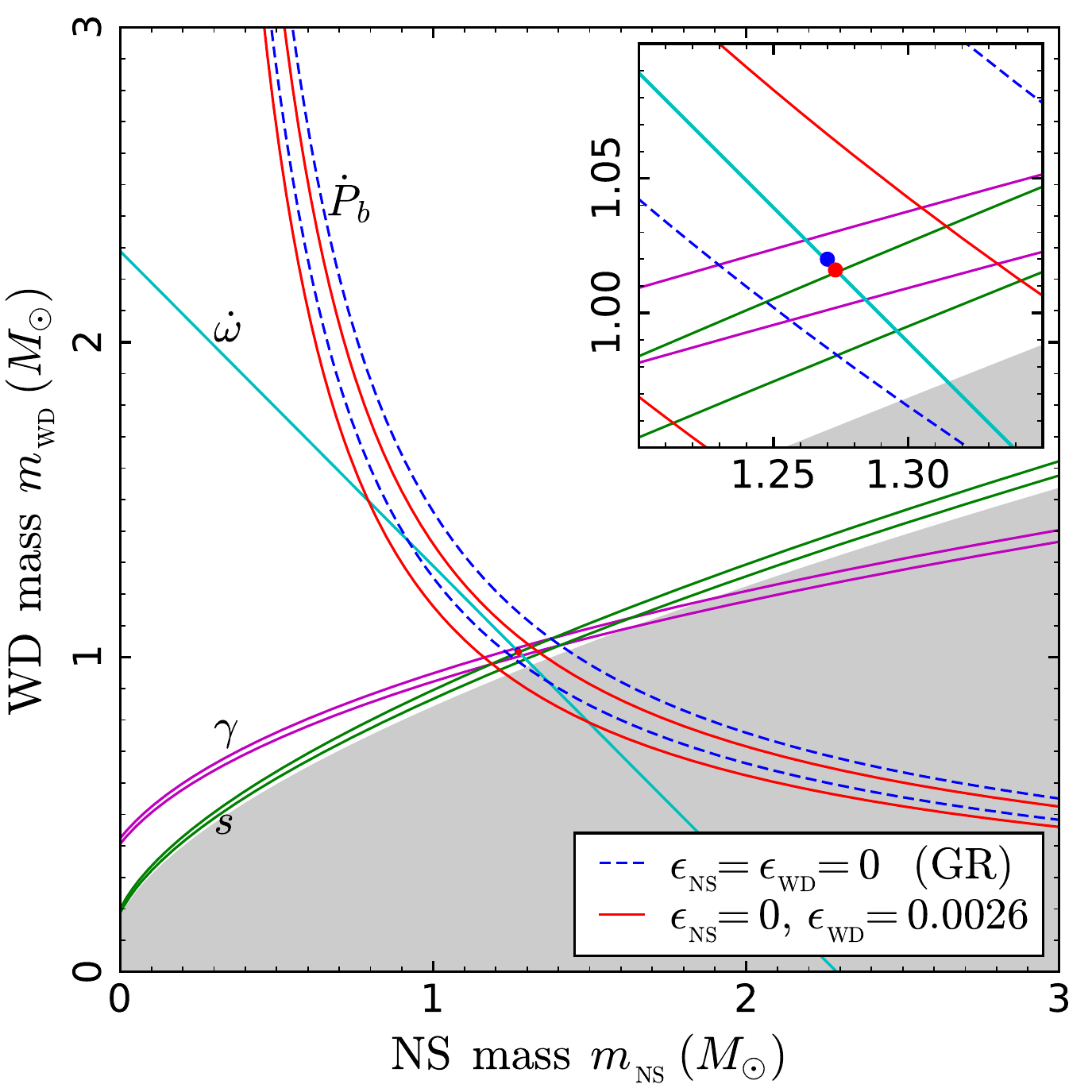}
\caption{Mass-mass diagram for PSR J1141$-$6545 based on the PK parameters: $\dot{\omega}$ (cyan), ${\gamma}$ (purple), $s$ (green), $\dot{P}_b$ in GR (blue dashed) and $\dot{P}_b$ in SMG (red). The separation of the curves represents $\pm1\sigma$ error bounds. The gray region is excluded by the condition $\sin i\le1$.
}
\label{fig1}
\end{figure}

We construct the mass-mass diagram in Fig. \ref{fig1}, and slight difference between the GR masses (blue dot) and SMG masses (red dot) is observed in the diagram. Since $\ep_{\N}=0$ is adopted, the PK parameters $\dot{\omega}$, $\gamma$ and $s$ in SMG are the same as those in GR, and labeled by cyan, purple and green, respectively. The blue dashed curves denote $\dot{P}_b$ in GR, and the red curves denote $\dot{P}_b$ in SMG, which is obtained by pushing $\ep_{\W}$ to its median value (0.0026). In general, these curves are different for different theories of gravity, but they should intersect in the same region if the theory is valid. Here, all curves intersect in the same region, meaning that SMG pass these tests.

%-----------------Fig1-----------------%
%-----------------Fig1-----------------%

%%%%%%%%%%%%%%%%%%%%%%%%%7
{\emph{Deviations from GR.}}---The screening mechanisms imply that the more compact the star is, the smaller its scalar charge is. In general, the compactness increases as the mass increases for low-mass WDs \cite{Panei2000p970}. Therefore, the scalar charge decreases as the mass increases, which is expected to be observed from WDs.

\begin{figure}[htbp]
\centering
\includegraphics[width=8cm, height=5.5cm]{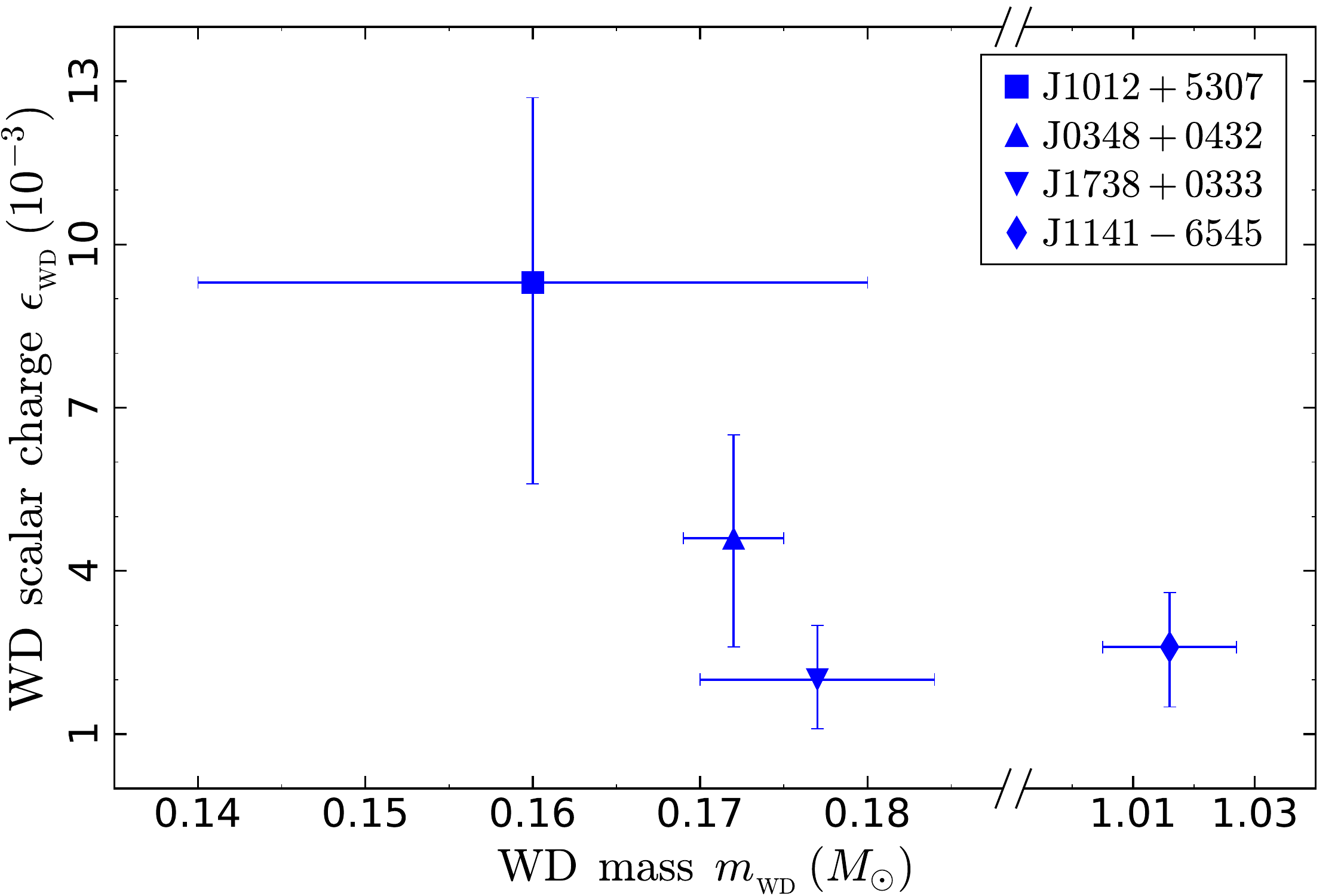}
\caption{Mass-scalar charge diagram for WDs from four independent systems. The error bars denote the 68.3\% CL.
}
\label{fig2}
\end{figure}

Fig. \ref{fig2} presents the mass-scalar charge diagram for WDs derived from the four independent systems under considerations. Indeed, we find that there exists a tendency of inverse correlation between $m_{\W}$ and $\epsilon_{\W}$ for three low-mass WDs (expect for the massive WD in PSR J1141$-$6545), which provides strong supports for the screening mechanisms. In addition, we find that the scalar charges of all WDs are not only of the same order of magnitude ($\sim10^{-3}$) but also inconsistent with zero at 95.4\%~CL (see Table \ref{tab2_results}), which represents slight but significant deviations from GR, although within the error bars our results are still marginally consistent with GR. Among them, PSR~J1738$+$0333 provides the most stringent constraint on the scalar charge, which is consistent with the conclusion given in \cite{Freire2012p3328a}.

%-----------------Fig2-----------------%
%-----------------Fig2-----------------%

Moreover, for the two WDs from PSRs J1738$+$0333 and J0348$+$0432, their radii are measured  (see Table \ref{tab1_PSRs}), hence their compactnesses are $\Phi_{\W}=1.04\times10^{-5}$ and $\Phi_{\W}=5.61\times10^{-6}$, respectively. Using Eq. \eqref{epsilon_a}, we obtain the constraints on the scalar field VEV (95.4\%~CL),
\begin{align}
\label{constr_phi}
\f{\ph}{\Mpl}=\left\{
\begin{matrix}
2.05^{+1.99}_{-1.65}\times10^{-8}~~~\rm~from~~~J1738+0333
\\[0.5 em]
2.56^{+1.99}_{-2.01}\times10^{-8}~~~\rm~from~~~J0348+0432
\end{matrix}
\right..
\end{align}
These two independent constraints coincide with each other, which provides a strong support for the relation \eqref{epsilon_a}. Combining them, we obtain
\begin{align}
\label{constr_2phi}
\f{\ph}{\Mpl}=2.30^{+2.07}_{-1.83}\times10^{-8}~~~~{\rm (95.4\%~CL)}.
\end{align}
As one of the main conclusions of this Letter, this result is applicable for any SMG model.

%%%%%%%%%%%%%%%%%%%%%%%%%8
{\emph{Application to chameleon.}}---The above results are generically applicable to all SMG theories, and here we consider the chameleon model as an example. The chameleon model was introduced as a screening mechanism by Khoury and Weltman \cite{Khoury2004p44026,Khoury2004p171104}. The original model is ruled out by combining the observations of the Solar System and cosmology \cite{Hees2012p103005,Zhang2016p124003}. We consider the exponential chameleon, which is characterized by an exponential potential and an exponential coupling function \cite{Brax2004p123518},
\begin{align}
\label{Chameleon_V_A}
V(\phi)=\Lambda^4\exp\Big(\frac{\Lambda^{\alpha}}{\phi^{\alpha}}\Big),
\quad~
A(\phi)=\exp\Big(\frac{\beta\phi}{\Mpl}\Big),
\end{align}
where $\alpha$ and $\beta$ are all positive constants, and $\Lambda$ labels the energy scale of the theory. The chameleon VEV is given by \cite{Zhang2017p104027}
\begin{align}
\ph=\bigg(\frac{{\alpha} \Mpl \Lambda^{4+\alpha}}{\beta\rho_{b}}\bigg)^{\frac{1}{\alpha+1}}
\label{chameleon_VEV},
\end{align}
where $\rho_{b}\simeq10^{-42}\,\rm GeV^4$ is the galactic background density.

%-----------------Fig3-----------------%
\begin{figure}[htbp]
\begin{center}
\includegraphics[width=8cm, height=5.5cm]{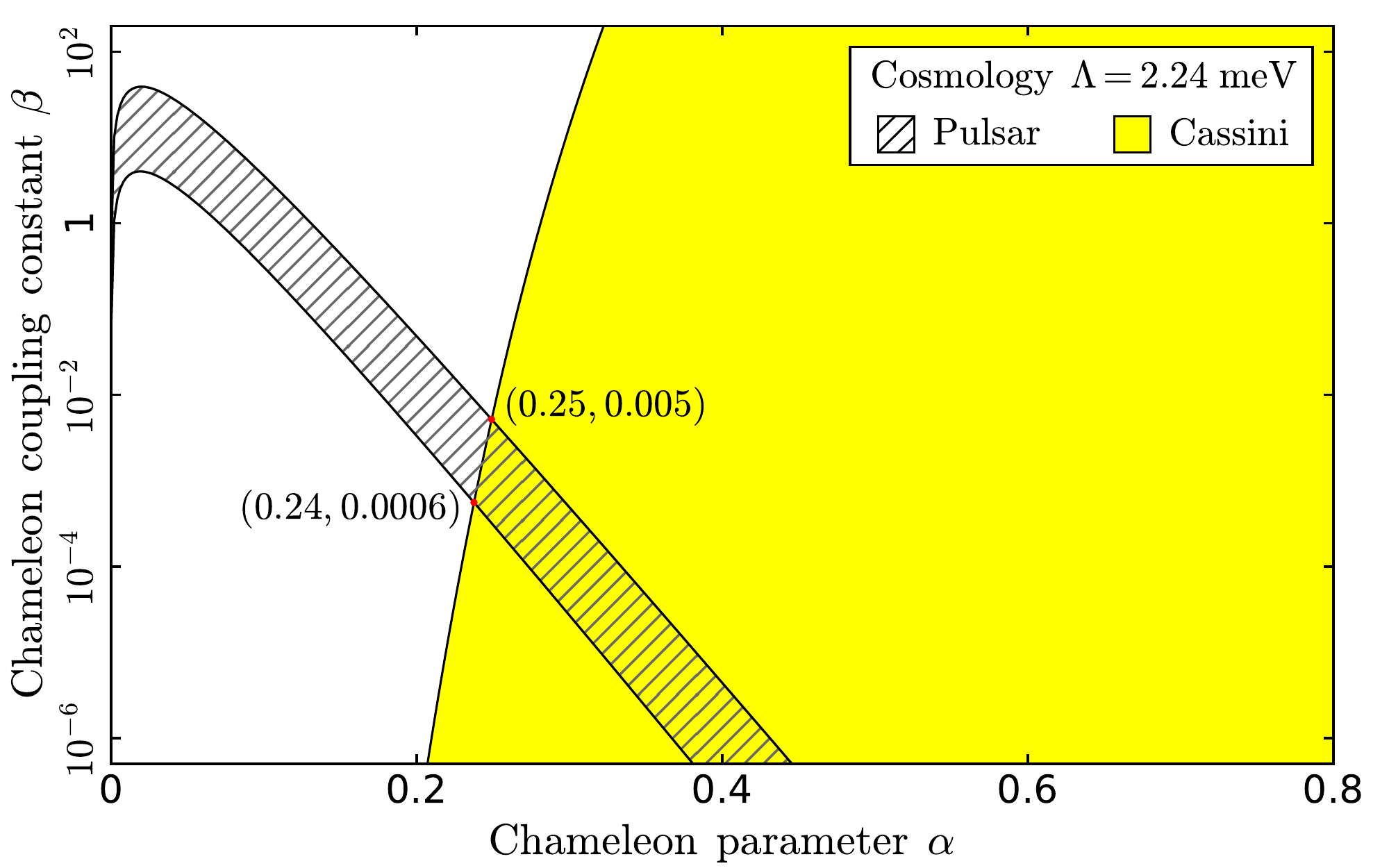}
\caption{In the chameleon parameter space, the yellow region is allowed by the tests of Cassini experiment, while the shadow region is allowed by the pulsar observations.
}
\label{fig3}
\end{center}
\end{figure}
%-----------------Fig3-----------------%

The cosmological constraints require that the energy scale $\Lambda$ is close to the dark energy scale, i.e. $\Lambda\simeq 2.24\times10^{-3}$\,eV \cite{Hamilton2015p849,Zhang2016p124003}. Using Eq. \eqref{chameleon_VEV}, from the pulsar constraint \eqref{constr_2phi}, the allowed parameter space of $(\alpha, \beta)$ is illustrated by the shadow region in Fig.~\ref{fig3}. The post-Newtonian parameter now is given by $\gamma_{\scriptscriptstyle\rm PPN}=1-2\beta\ph/(\Mpl\Phi)$ \cite{Zhang2016p124003}. Then, from the Cassini constraint $\left|\gamma_{\scriptscriptstyle\rm PPN}-1\right|\le2.3\times10^{-5}$ \cite{Bertotti2003p374}, the allowed parameter space of $(\alpha, \beta)$ is shown by the yellow region in Fig. \ref{fig3}. The overlap region yields a lower bound $\alpha\ge0.24$ and an upper bound $\beta\le0.005$, which implies that the chameleon is weakly coupled to matter.

%%%%%%%%%%%%%%%%%%%%%%%%%9
{\emph{Conclusions.}}---As a simple generalization of GR, SMG is a class of scalar-tensor theories of gravity with screening mechanisms in order to satisfy the tight Solar System tests. In this Letter, we have constrained the general SMG by using the full observed PK parameters of NS-WD binary pulsar systems, in which the difference from GR is characterized by a unique parameter, i.e., the scalar charge  $\epsilon_{\W}$ of WD. Considering four independent binaries, we have found the nonzero values of $\epsilon_{\W}$ for all WDs, and also an anticorrelation between $\epsilon_{\W}$ and $m_{\W}$ in the low-mass WDs, which is consistent with the prediction of the screening mechanisms. Especially, from two different binaries (PSRs J1738$+$0333 and J0348$+$0432) with measured radii of WDs, we have obtained the coincident constraints on the scalar field VEV. These self-consistent results show explicitly the deviations from GR in binary pulsars at the 2$\sigma$ level. To confirm our results, further observations with more precise measurements are highly demanded.

~

%%%%%%%%%%%%%%%%%%%%%%%%%
%{\emph{Acknowledgments}.---
This work is supported in part by the National Natural Foundation of China (NNSFC) with the grant numbers:  Nos. 11603020, 11633001, 11173021, 11322324, 11653002, 11421303, 11375153, 11675145, 11675143, and 11105120.

%\bibliographystyle{prsty3-1.bst}
%\bibliography{/Users/starzhx/Desktop/bib/starzhx}

\begin{thebibliography}{40}

\bibitem{Will1993p}
C.~M. Will, {\em Theory and Experiment in Gravitational Physics} (Cambridge
  University Press, Cambridge, England, 1993).

\bibitem{Popper2002p}
K. Popper, {\em The Logic of Scientific Discovery}, {\em 2nd English ed.}
  (Routledge Press, London and New York, 2002).

\bibitem{Fujii2003p}
Y. Fujii and K.~I. Maeda, {\em The Scalar-Tensor Theory of Gravitation}
  (Cambridge University Press, Cambridge, 2003).

\bibitem{Damour1992p2093}
T. Damour and G. Esposito-Farese, Class. Quant. Grav. {\bf 9},  2093  (1992).

\bibitem{Khoury2004p44026}
J. Khoury and A. Weltman, \prd {\bf 69},  044026  (2004).

\bibitem{Khoury2004p171104}
J. Khoury and A. Weltman, Phys. Rev. Lett. {\bf 93},  171104  (2004).

\bibitem{Hinterbichler2010p231301}
K. Hinterbichler and J. Khoury, Phys. Rev. Lett. {\bf 104},  231301  (2010).

\bibitem{Hinterbichler2011p103521}
K. {Hinterbichler}, J. {Khoury}, A. {Levy}, and A. {Matas}, \prd {\bf 84},
  103521  (2011).

\bibitem{Damour1994p532}
T. {Damour} and A.~M. {Polyakov}, Nuclear Physics B {\bf 423},  532  (1994).

\bibitem{Brax2010p63519}
P. Brax, C. van~de Bruck, A.-C. Davis, and D. Shaw, \prd {\bf 82},
  063519  (2010).

\bibitem{Brax2012p44015}
P. Brax, A.-C. Davis, B. Li, and H.~A. Winther, \prd {\bf 86},  044015
  (2012).

\bibitem{stairs2003}
I. H. Stairs, Living Rev. Relativ. {\bf 6}, 5 (2003).

\bibitem{ST-pulsar}
T. Damour and G. Esposito-Farese, \prd {\bf 54}, 1474 (1996); {\bf 58}, 042001 (1998);
T. Damour, arXiv:0704.0749.

\bibitem{other-pulsar}
K. Yagi, D. Blas, N.  Yunes and  E. Barausse, Phys. Rev. Lett. {\bf 112}, 161101 (2014);
J. B. Jimenez, F. Piazza and H. Velten, Phys. Rev. Lett. {\bf 116}, 061101 (2016).

\bibitem{Zhang2017p104027}
X. {Zhang}, T. {Liu}, and W. {Zhao}, \prd {\bf 95},  104027  (2017).

\bibitem{Gronke2015p123}
M. Gronke, D.~F. Mota, and H.~A. Winther, Astron. Astrophys. {\bf 583},  A123
  (2015).

\bibitem{Schmidt2010p103002}
F. {Schmidt}, \prd {\bf 81},  103002  (2010).

\bibitem{liu2018}
T. Liu, X. Zhang, W. Zhao et al., in preparation.

\bibitem{Abbott2017p13}
B.~P. Abbott {\it et~al.}, Astrophys. J. {\bf 848},  L13  (2017).

\bibitem{Abbott2017p161101}
B. Abbott {\it et~al.}, Phys. Rev. Lett. {\bf 119},  161101  (2017).

\bibitem{Damour1986p263}
T. {Damour} and N. {Deruelle}, Ann.~Inst.~Henri Poincar{\'e} Phys.~Th{\'e}or., {\bf 44},  263  (1986).

\bibitem{Damour1992p1840}
T. {Damour} and J.~H. {Taylor}, \prd {\bf 45},  1840  (1992).

\bibitem{Zhangp}
X. Zhang, W. Zhao, T. Liu, K. Lin, A. Wang, C. Zhang, S. Zhang, X. Zhao, and T. Zhu, in preparation.

\bibitem{Eardley1975p59a}
D.~M. {Eardley}, Astrophys. J. {\bf 196},  L59  (1975).

\bibitem{Zhang2016p124003}
X. Zhang, W. Zhao, H. Huang, and Y. Cai, \prd {\bf 93},  124003  (2016).

\bibitem{footnote1}
Note that the observed value of $\dot{P}_b$ needs to be corrected by two main effects, the differential galactic acceleration \cite{Damour1991p501} and the Shklovskii effect \cite{Shklovskii1970p562}, to obtain its intrinsic value caused by gravitational radiation damping.

\bibitem{Damour1991p501}
T. {Damour} and J.~H. {Taylor}, Astrophys. J. {\bf 366},  501  (1991).

\bibitem{Shklovskii1970p562}
I.~S. {Shklovskii}, Sov.~Astron. {\bf 13},  562  (1970).

\bibitem{Bhat2008p124017}
N.~D.~R. Bhat, M. Bailes, and J.~P.~W. Verbiest, \prd {\bf 77},  124017
  (2008).

\bibitem{Ord2002p75}
S.~M. Ord, M. Bailes, and W. van Straten, Astrophys. J. {\bf 574},  L75
  (2002).

\bibitem{Freire2012p3328a}
P.~C.~C. Freire {et~al.}, Mon. Not. Roy. Astron. Soc. {\bf 423},  3328
  (2012).

\bibitem{Antoniadis2013p6131}
J. Antoniadis {et~al.}, Science {\bf 340},  6131  (2013).

\bibitem{Lazaridis2009p805}
K. Lazaridis {et~al.}, Mon. Not. R. Astron. Soc. {\bf 400},  805  (2009).

\bibitem{Desvignes2016p3341a}
G. Desvignes {et~al.}, Mon. Not. Roy. Astron. Soc. {\bf 458},  3341
  (2016).

\bibitem{Panei2000p970}
J.~A. {Panei}, L.~G. {Althaus}, and O.~G. {Benvenuto}, A\&A {\bf 353},  970
  (2000).

\bibitem{Hees2012p103005}
A. {Hees} and A. {F{\"u}zfa}, \prd {\bf 85},  103005  (2012).

\bibitem{Brax2004p123518}
P. {Brax} {et~al.}, \prd {\bf 70},  123518  (2004).

\bibitem{Hamilton2015p849}
P. {Hamilton} {et~al.}, Science {\bf 349},  849  (2015).

\bibitem{Bertotti2003p374}
B. Bertotti, L. Iess, and P. Tortora, Nature {\bf 425},  374  (2003).


\end{thebibliography}

\end{document}